\documentstyle[12pt]{article}
\title{\sf GROUND STATE PROPERTIES OF EXOTIC NUCLEI NEAR Z=40
IN THE RELATIVISTIC MEAN-FIELD THEORY}
\author{G.A.  LALAZISSIS$^1$ and M.M. SHARMA$^2$ \\
$^1$Physik Department, Technische Universit\"at M\"unchen \\
D-85747 Garching, Germany\\
$^2$Max Planck Institut f\"ur Astrophysik, Karl-Schwarzschild-Strasse 1,\\
D-85740 Garching bei M\"unchen, Germany}

\begin{document}
\maketitle
\begin{abstract}
\baselineskip 18pt
Study of the ground-state properties of Kr, Sr and Zr isotopes has
been performed in the framework of the relativistic mean field (RMF)
theory using the recently proposed relativistic parameter set NL-SH.
It is shown that the RMF theory provides an unified and excellent
description of the binding energies, isotope shifts and deformation
properties of nuclei over a large range of isospin in the Z=40 region.
It is observed that the RMF theory with the force NL-SH is able to
describe the anomalous kinks in isotope shifts in Kr and Sr nuclei,
the problem which has hitherto remained unresolved. This is in contrast
with the density-dependent Skyrme Hartree-Fock approach which does not
reproduce the behaviour of the isotope shifts about shell closure.
On the Zr chain we predict that the isotope shifts exhibit a trend
similar to that of the Kr and Sr nuclei. The RMF theory also predicts
shape coexistence in heavy Sr isotopes. Several dramatic shape transitions
in the isotopic chains are shown to be a general feature of nuclei in
this region. A comparison of the properties with the available
mass models shows that the results of the RMF theory are generally in
accord with the predictions of the finite-range droplet model.
\end{abstract}

\newpage
\baselineskip 18 pt
\section{\sf INTRODUCTION}

The relativistic mean-field (RMF) theory \cite{ser86} has received much
attention recently due to its advantages over the non-relativistic
density-depdendent Skyrme approaches \cite{brin72}. The RMF theory
has been successful \cite{rein89,gam90} in describing the
ground-state properties of nuclei about the line of stability.
It has been demonstrated \cite{sha93} that the RMF theory is able
to describe nuclei also far away from the line of stability.
With perpetual improvement in the techniques of producing radioactive
beams, study of exotic nuclei \cite{muel93,boyd94} very far
away from the stability line has now become feasible. Interesting
features such as neutron halos \cite{tani} in neutron-rich nuclei
are being unravelled. As nuclei very far away from the stability
line entail very large isospins, it is incumbent on theories to be
able to provide and predict the properties of nuclei in these extreme
regions. Microscopic theories which could fulfill these demands are
much required. Availability of appropriate interactions to be able to
do justice to these extreme conditions has been equally in demand.
The RMF theory bears a potential to describing nuclei over a large
range of charge and mass.

The RMF theory has been applied extensively to study the ground-state
properties of nuclei along the stablity line. The forces NL1 \cite{rein89}
and NL2 \cite{lee86} were employed in most of the calculations \cite{gam90}.
NL1 and NL2 were the only available choice for realistic calculations.
It was, however, observed \cite{sha92} that both these parameter sets
overestimate the neutron-skin thickness of nuclei with large neutron
excess. A larger asymmetry energy of both NL1 and NL2 was found responsible
for this problem. Recently, this problem has been remedied and a new
force NL-SH was obtained by Sharma et al. \cite{sha93}. This force has
been shown to provide an excellent description of the ground state
properties of spherical as well as deformed nuclei along the stability
line. The description obtained for nuclei away from the stability line
\cite{sha93} and close to the neutron drip line \cite{sha94} has also
been successful. This is partly due to the fact that the force NL-SH has
an appropriate $\rho$-meson coupling  which leads to the correct asymmetry
energy. It was shown \cite{sha93} that NL-SH, which was obtained by
adjusting the ground-state properties of only a few spherical nuclei,
describes the binding energies and deformations of Xe isotopes over a
large range of isopin on both the sides of the stability line.

The density-dependent Skyrme approaches \cite{brin72} have, on the other
hand, also been applied successfully to describing properties of nuclei
mainly along the stability line. Properties of deformed nuclei have been
predicted \cite{quen} satisfactorily using the Skyrme force SIII.
The Skyrme approach, however, faces problems in providing an adequate
description of nuclei across shell-closures. It is worth mentioning
that the two approaches, one that of the RMF theory and the other of
the Skyrme type have quite different underlying assumptions to generate
the interaction. In the RMF theory the saturation and the density
dependence of the nuclear interaction is provided by a balance between
a large attractive scalar $\sigma$-meson field and a large repulsive
vector $\omega$-meson field. The asymmetry component is provided by the
isovector $\rho$ meson. The nuclear interaction is hence generated by
the exchange of various mesons between nucleons. On the other hand,
in the Skyrme approach, the nuclear interaction has a certain density
dependence assumed at the outset, the latter being obtained from fits
to empirical data. It may be reckoned that the structure of the force
in the RMF theory and its density dependence differ from that of the
Skyrme forces. Moreover, the spin-orbit interaction is added ad-hoc
in the latter approach. It, however, arises naturally in the RMF theory
as a result of the Dirac structure of nucleons. The density dependence of
the spin-orbit interaction in the Skyrme theory also differs from its
counterpart in the RMF theory. We have found that the isospin dependence
of the spin-orbit interaction in the two approaches is different
\cite{lala94,koen94}.

The isotope shifts of nuclei have been measured since long. Precision
measurements on isotope shifts of nuclei have been carried out
extensively by the Mainz and other groups on several chains of nuclei.
A detailed discussion of these properties has been provided in
a review article by Otten \cite{ott88}. As shown in this review,
many isotopic chains exhibit anomalous kinks in the isotope shifts
about neutron magic number. Such chains include the Pb isotopes
which also have a closed proton shell. Other chains which also show such
a behaviour include the isotopes of Sr and Kr among others. These
isotopes do not have a magic proton number but encounter a closed
neutron shell at N=50. The anomalous behaviour of the charge radii of
nuclei about the shell-closure has been one of the long-standing
problems. The empirical data which have reached considerable precision
using laser spectroscopic methods show this behaviour conspicuoulsy.
A kink in the isotope shifts in charge radii implies that the charge
radii of the isotopes above the closed shell have a different trend
in its variation with the neutron number as compared with their
lighter counterparts. This suggests that neutrons in the unfilled-shell
below the shell-closure behave differently  than those
which fill the next shell above it.

The problem of the anomalous kink in the empirical isotope shifts
of Pb nuclei has until recently remained intractable. This problem was
treated earlier using the Skyrme approach. It was shown by Sagawa et al.
\cite{sag87} that the isotope shifts only on one side of the neutron
closed shell could be reproduced. A detailed investigation on it was
also undertaken by Tajima et al. \cite{taji93}. Employing several Skyrme
interactions, it was shown \cite{taji93} that the kink in the isotope
shifts in Pb nuclei about N=126 could not be obtained within the
Skyrme mean-field. Adding various correlations to the mean-field
improved the situation only slightly. The kink in the isotope shifts
could not, however, be reproduced by any of the Skyrme forces used
in \cite{taji93}.  Recently, an attempt was made by Sharma et al.
\cite{sha93b} to obtain the isotope shifts in Pb nuclei in the RMF
theory. It was shown that the RMF theory with the force NL-SH is
successful in reproducing the anomalous kink in the isotope shifts of
Pb nuclei.

Nuclei in the region of Z=40 about the mass number 100 have been of particular
interest in the nuclear structure due to predictions for the existence
of highly deformed shapes. Properties of nuclei in this region have in
the past been the focus of several detailed investigations both theoretically
and experimentally. Extensive efforts devoted to studies in this region
have been discussed in detail in \cite{conf}. On the experimental front,
there exist high precision data on isotope shifts for Kr and Sr nuclei
obtained by the Mainz group \cite{buch90,klein94} using laser spectroscopic
methods. Measurements on the Sr chain were also carried out at the
Daresbury Laboratory \cite{east87}. The data show an interesting feature
that the magic nuclei $^{86}$Kr and $^{88}$Sr have the smallest rms charge
radii in their respective chains, even when compared to nuclei ten mass
numbers lower.

Theoretically, the isotope shifts and charge radii of nuclei in this
region are still beyond a description. Extensive investigations
were carried out by Bonche et al. \cite{bonche85}. Self-consistent
calculations \cite{bonche85} including triaxial deformations in the
Skyrme mean-field with the force SIII provide only a nearly linear
response to the isotope shifts with mass number, in clear contradiction
to the experimental trend. Furthermore, application of the method of
generator coordinates \cite{bonche91,heen93} to include triaxial
deformations in Sr isotopes predicts spherical to deformed shape
transitions. The isotope shifts could not, however, be reproduced by
any means in the non-relativistic Skyrme approaches.

Attention has also been paid to the Sr and Zr isotopes
\cite{maha92,shei93,hira93} in the RMF theory. However, because of the
use of the force NL1 \cite{rein89}, which has an asymmetry energy much
larger than the empirical value, a good description for isotopes with
a large neutron excess has been hindered. The RMF theory with NL1 has
shown success primarily for nuclei along the line of the stability
\cite{gam90}. Due to inadequacy of the force NL1 to extrapolations
in the extreme regions, it has predicted spherical shapes for many
Sr \cite{maha92} as well as Zr isotopes \cite{shei93}, in contradiction
to the deformed shapes obtained empirically. With the advent of the
force NL-SH, we study the ground state properties of Kr, Sr and Zr
isotopes over a large range of isospin within the framework of the
RMF theory in this paper. We focus especially upon the isotope shifts
and deformation properties of these nuclei. The paper is organized
as follows. Section II describes the details of the theoretical formalism
employed in the RMF treatment of the deformed nuclei. Section III gives
details on the procedure of our calculations. In section IV we present
and discuss our results, where a comparison of our predictions with the
empirical data is made. Comparison with other approaches is also made
wherevere possible. The last section summarizes our conclusions.

\section{\sf THE RELATIVISTIC MEAN-FIELD FORMALISM}

The starting point of the RMF theory is a Lagrangian density \cite{ser86}
which describes the nucleons as Dirac spinors interacting via the
exchange of several mesons. The Lagrangian density can be written in
the following form:
\begin{equation}
\begin{array}{rl}
{\cal L} &=
\bar \psi (i\rlap{/}\partial -M) \psi +
\,{1\over2}\partial_\mu\sigma\partial^\mu\sigma-U(\sigma)
-{1\over4}\Omega_{\mu\nu}\Omega^{\mu\nu}+\\
\                                        \\
\ & {1\over2}m_\omega^2\omega_\mu\omega^\mu
-{1\over4}{\bf R}_{\mu\nu}{\bf R}^{\mu\nu} +
 {1\over2}m_{\rho}^{2}
 \mbox{\boldmath $\rho$}_{\mu}\mbox{\boldmath $\rho$}^{\mu}
-{1\over4}F_{\mu\nu}F^{\mu\nu} \\
\                              \\
\ &  g_{\sigma}\bar\psi \sigma \psi~
     -~g_{\omega}\bar\psi \rlap{/}\omega \psi~
     -~g_{\rho}  \bar\psi
      \rlap{/}\mbox{\boldmath $\rho$}
      \mbox{\boldmath $\tau$} \psi
     -~e \bar\psi \rlap{/}{\bf A} \psi
\end{array}
\end{equation}

TBhe meson fields are the isoscalar $\sigma$ meson, the isoscalar-vector
$\omega$ meson and the isovector-vector $\rho$ meson. The latter provides
the necessary isospin asymmetry. The bold-faced letters indicate the
isovector quantities. The model contains also a non-linear scalar
self-interaction of the $\sigma$ meson :

\begin{equation}
U(\sigma)~={1\over2}m_{\sigma}^{2} \sigma^{2}~+~
{1\over3}g_{2}\sigma^{3}~+~{1\over4}g_{3}\sigma^{4}
\end{equation}
The scalar potential (2) introduced by Boguta and Bodmer \cite{bog77} is
essential for appropriate description of surface properties.
M, m$_{\sigma}$, m$_{\omega}$ and m$_{\rho}$ are the nucleon-, the $\sigma$-,
the $\omega$- and the $\rho$-meson masses respectively, while g$_{\sigma}$,
g$_{\omega}$, g$_{\rho}$ and e$^2$/4$\pi$ = 1/137 are the corresponding
coupling constants for the mesons and the photon.

The field tensors of the vector mesons and of the electromagnetic
field take the following form:
\begin{equation}
\begin {array}{rl}
\Omega^{\mu\nu} =& \partial^{\mu}\omega^{\nu}-\partial^{\nu}\omega^{\mu}\\
\          \\
{\bf R}^{\mu\nu} =& \partial^{\mu}
                  \mbox{\boldmath $\rho$}^{\nu}
                  -\partial^{\nu}
                  \mbox{\boldmath $\rho$}^{\mu}
                  -g_{\rho}(
                  \mbox{\boldmath $\rho$} \times
                  \mbox{\boldmath $\rho$})\\
\                \\
F^{\mu\nu} =& \partial^{\mu}{\bf A}^{\nu}-\partial^{\nu}{\bf A}^{\mu}
\end{array}
\end{equation}
The classical variational principle gives the equations of motion. In our
approach, where time reversal and charge conservation is considered, the
Dirac equation is written as:
\begin{equation}
\{ -i{\bf {\alpha}} \nabla + V({\bf r}) + \beta [ M + S({\bf r}) ] \}
\end{equation}
where $V({\bf r})$ represents the $vector$ potential:
\begin{equation}
V({\bf r}) = g_{\omega} \omega_{0}({\bf r}) + g_{\rho}\tau_{3} {\bf {\rho}}
_{0}({\bf r}) + e{1+\tau_{3} \over 2} {\bf A}_{0}({\bf r})
\end{equation}
and $S({\bf r})$ the $scalar$ potential:
\begin{equation}
S({\bf r}) = g_{\sigma} \sigma({\bf r})
\end{equation}
the latter contributes to the effective mass as:
\begin{equation}
M^{\ast}({\bf r}) = M + S({\bf r})
\end{equation}
The Klein-Gordon equations for the meson fields are time-independent
inhomogenous equations with the nucleon densities as sources.
\begin{equation}
\begin{array}{ll}
\{ -\Delta + m_{\sigma}^{2} \}\sigma({\bf r})
 =& -g_{\sigma}\rho_{s}({\bf r})
-g_{2}\sigma^{2}({\bf r})-g_{3}^{3}({\bf r})\\
\         \\
\  \{ -\Delta + m_{\omega}^{2} \} \omega_{0}({\bf r})
=& g_{\omega}\rho_{v}({\bf r})\\
\                            \\
\  \{ -\Delta + m_{\rho}^{2} \}\rho_{0}({\bf r})
=& g_{\rho} \rho_{3}({\bf r})\\
\                           \\
\  -\Delta A_{0}({\bf r}) = e\rho_{c}({\bf r})
\end{array}
\end{equation}
The corresponding source terms are
\begin{equation}
\begin{array}{ll}
\rho_{s} =& \sum\limits_{i=1}^{A} \bar\psi_{i}~\psi_{i}\\
\             \\
\rho_{v} =& \sum\limits_{i=1}^{A} \psi^{+}_{i}~\psi_{i}\\
\             \\
\rho_{3} =& \sum\limits_{p=1}^{Z}\psi^{+}_{p}~\psi_{p}~-~
\sum\limits_{n=1}^{N} \psi^{+}_{n}~\psi_{n}\\
\                    \\
\ \rho_{c} =& \sum\limits_{p=1}^{Z} \psi^{+}_{p}~\psi_{p}
\end{array}
\end{equation}
where the sums are taken over the valence nucleons only. It should also
be noted that the present approach neglects the contributions of
negative-energy states ($no-sea$ approximation), i.e. the vacuum is not
polarized.

The Dirac equation is solved using the oscillator expansion method
\cite{gam90}. For the determination of the basis wavefunctions an
axially symmetric harmonic oscillator potential with size parameters
\begin{equation}
b_{z} = b_{z}(b_{0},\beta_{0})= \sqrt{\hbar/M\omega_{z}}~~~~~~~~~~
b_{\bot}=b_{\bot}(b_{0},\beta_{0})=
 \sqrt{\hbar/M\omega_{\bot}}
\end{equation}
is employed. The basis is defined in terms of $b_{0}$ and the deformation
parameter $\beta_{0}$. The volume conservation relates the quantities
$b_{z}$, $b_{\bot}$ and $b_{0}$ by $b_{\bot}^{2}b_{z}=b_{0}^{3}$.

\section {\sf DETAILS OF THE CALCULATIONS}

We have performed relativistic Hartree calculations for three
isotopic chains. The isotopes of Kr (Z=36), Sr (Z=38) and Zr
(Z=40) have been considered here. As many nuclei in these chains
are well deformed and involve several shape transitions,
the calculations have been carried out for an axially deformed
configuration for all the nuclei. Calculations for even Sr isotopes
encompass mass numbers A=70 up to A=110, while the Kr isotopes cover
the region between A=72 and A=100. We have also performed calculations
for Zr nuclei in the region 80$\leq$A$\leq$110. On the isotopic chains
of Sr and Kr, several precision measurements \cite{ott88,buch90,klein94}
on isotopic shifts are available. For Zr isotopes, where measurements do
not yet exist for the whole chain, our predictions serve to illustrate the
comparison of the trend of the isotope shifts in this mass region.

For open shell nuclei, pairing has been included using the BCS formalism.
In the BCS calculations we have used constant pairing gaps,
which are taken from the empirical particle separation energies
of neighbouring nuclei. The zero-point energy of a harmonic oscillator
for the centre-of-mass energy is also included. In the present approach
angular momentum and particle number projection as well as collective
vibrations are neglected. It is estimated, however, that these additional
features will have very small contributions.

The number of shells taken into account is 12 for both fermionic and
bosonic wavefunctions. It should be noted that for convergence reasons
14 shells were also considered. It turned out, however, that
the difference in the results is negligible and therefore all the
calculations reported in the present work were performed in a
12 shells harmonic oscillator expansion.

In this paper we have used the recently proposed force NL-SH.
This force has been shown to provide excellent results \cite{sha93}
for nuclei on both the sides of the stability line. The good value of
the asymmetry energy of this force is partly responsible for a proper
description of nuclei on both the proton rich and neutron rich sides.
The parameters of the force NL-SH were provided in \cite{sha93}.
However, rounding off of some parameters to 3 decimal places as given
in \cite{sha93} overestimates the binding energies by a few MeV.
The exact values of the force parameters were given in \cite{sha93b}.
The parameters of the force NL-SH are:

\par\noindent
M = 939.0 MeV; $m_\sigma$ = 526.059 MeV; $m_\omega$ = 783.0 MeV;
$m_\rho$ = 763.0 MeV;
\par\noindent
$g_\sigma$ = 10.444; $g_\omega$ = 12.945; $g_\rho$ = 4.383;
$g_2$ = $-$6.9099 fm$^{-1}$ ; $g_3$ = $-$15.8337.

Here $g_2$ is in fm$^{-1}$ and $g_3$ is dimensionless.
It should be pointed out that in refs.\cite{sha93,sha93b}
the dimensions of $g_2$ and $g_3$ were exchanged
by mistake.

\bigskip
\section {\sf RESULTS AND DISCUSSION}

\subsection{\sf Binding Energy}
\bigskip
\noindent
In Fig. 1 we show the binding energy per nucleon (E/A) for
Kr, Sr and Zr nuclei using the force NL-SH in the RMF theory.
The empirical values taken from the 1993 Atomic Mass Evaluation
Tables \cite{audi93} (expt.) are also shown. The figure
also includes the predictions of the recent finite-range droplet
model \cite{nix94} (FRDM) and of the extended Thomas-Fermi with Strutinsky
Integral (ETF-SI) model \cite{abou92,pear93} for comparison. The parabolic
shapes of the binding energy per nucleon in all the three isotopic
chains is depicted very well in the figure. The minimum in the binding
energy is observed at the magic neutron number N=50 in the RMF as well
as in the mass models.
The RMF theory predicts binding energies which are in accord with
the empirical values in almost all the cases. In Tables
1-3 we present the total binding energies obtained with the force NL-SH.
For comparison, the empirical values \cite{audi93} are also shown.
The experimental binding energy of the nucleus $^{76}$Sr derives from
the recent mass measurement \cite{otto94} at the isotope on-line
separator ISOLDE at CERN. The predicted values from the mass formulae
FRDM and ETF-SI are also provided in the tables. We also show the values
obtained with the RMF force NL1 and with the Skyrme force SIII \cite{bonche85}
wherever available. In the latter case, triaxial deformations were also
taken into account. The SIII results underestimate the binding of most
of the nuclei by up to 6 MeV as one moves away from the closed
neutron-shell nuclei.

The binding energies in the RMF are in overall agreement with the
empirical data. Only for the light Kr isotopes does the RMF theory
provide the binding energies which are at the most 3-4 MeV less than
the empirical values. This disagreement amounts to 0.5\% at the maximum.
The empirical mass excess for many of the nuclei considered in our paper
are not known. In the RMF theory, we predict the binding energies of
these nuclei which lie mostly at the highly neutron deficient or
neutron-rich sides of the isotopic chains. The calculated total binding
energies in the RMF theory are also very close to those of FRDM and
ETFSI predictions within 1-2 MeV. That the FRDM and ETF-SI values are
close to the empirical ones in understandable, for these models have
been fitted exhaustively to a large number of known data. In contrast,
the agreement of the RMF theory with the empirical data and the
model values is remarkable notwithstanding the fact that the force
NL-SH was fitted only to 6 spherical nuclei at the stability line.

\bigskip
\subsection{\sf The Nuclear Radii and Isotope Shifts}
\bigskip
\noindent
In Fig. 2  we show the isotope shifts $r^{2}_{c}(A) - r^{2}_{c}(ref))$
for Kr, Sr and Zr nuclei calculated with respect to a reference nucleus
in each chain. The semi-magic nuclei $^{86}$Kr, $^{88}$Sr and
$^{90}$Zr serve as reference points. For the isotopic chains of
Kr \cite{klein94} and Sr \cite{ott88,buch90} the empirical data
obtained from atomic laser spectroscopy are also shown. The experimental
data for Kr and Sr nuclei exhibit a kink about the magic neutron number.
This kink about the closed-shell is a characteristic feature of
isotope shifts in many nuclei. This aspect has been clearly
illustrated in \cite{ott88}. A solution to this problem has eluded
since long. It can be seen from the figure that the RMF theory is successful
in reproducing this kink. In addition, the general trend which is
exhibited by the data is reproduced well. That is, the magic isotopes
$^{86}$Kr and $^{88}$Sr have the smallest rms charge radius in their
respective chains, even when compared to nuclei 10 mass numbers lower.
A similar behaviour is also predicted for Zr nuclei as can be seen from
Fig. 2 (c), where empirical isotope shifts on this chain of nuclei are not
known.

The absolute values of rms charge radii and neutron radii for all the three
isotopic chains are presented in Fig. 3. On going from the lighter isotopes
to the heavier ones, the charge radii exhibit a decreasing trend upto
thBe magic isotopes. Clearly the lighter nuclei below the magic numbers
possess higher charge radii than the heavier closed neutron-shell
nucleus. This is consistent with the upward kink in the isotope shifts of
lighter nuclei in Figure 2. The charge radii for nuclei heavier than the
clo[Bsed neutron-shell start increasing on adding further neutrons. The
neutron radii, on the other hand, also show a kink about the neutron shell
closure. However, the neutron radius for lighter isotopes in these chains
is not higher than that of the respective closed-shell nucleus.

The isotope shifts for the Zr chain have not been measured. Here we only
predict the behaviour of the isotope shifts in Zr nuclei. As seen in Fig. 2,
the isotope shifts in Zr nuclei also show a behaviour much similar to that
shown by Sr and Kr nuclei. For, isotope shifts are closely related to
shapes and deformation properties, all the three chains of nuclei seem
to exhibit a similar pattern in their spatial properties. We will discuss
the deformation properties in the ensuing subsection. Here we compare in
Fig. 4 our predictions for the isotope shifts of Zr nuclei with those
obtained with the Skyrme force SIII in \cite{bonche85}. The Skyrme approach
gives practically a straight line. A similar feature was also observed
\cite{taji93} for the Pb isotopes both with SkM* and SIII. Thus, the Skyrme
mean-field approach shows a trend very different from that of the empirical
data and the RMF theory.

In Figure 5 we show the neutron skin thickness $r_{n}-r_{p}$ for Zr
isotopes obtained in the RMF theory using the force NL-SH. We also
compare these results with those from NL1. It may be noted that the
force NL1 was used earlier in \cite{shei93} to calculate the properties
of neutron-rich Zr isotopes. Clear differences in the neutron-skin
thickness for NL-SH and NL1 can be seen with an increase in neutron
excess. This can be attributed primarily to the large asymmetry
energy of 44 MeV of NL1. The asymmetry energy of NL-SH is 36 MeV,
which is close to the empirical value of about 33 MeV.
The kink seen in the neutron-skin thickness for the force NL1
at A=100 arises due to the spherical shapes predicted by NL1 about
this mass.

In Fig. 5 we also show the neutron-skin thickness obtained
from the density-dependent Skyrme force SIII, taken from \cite{bonche85}.
For nuclei close to the symmetric nucleus $^{80}$Zr,
SIII gives neutron-skin thickness similar to those of NL-SH and NL1.
However, for nuclei with large neutron excess the values from SIII are
much smaller than those of NL-SH and NL1. This is due to the fact that
the force SIII provides neutron-rich Zr nuclei which have a larger
charge radius and a smaller neutron radius as compared to the RMF
forces.
\bigskip
\subsection{\sf Deformations}
\bigskip
\noindent
We have performed calculations in the RMF theory for both the prolate
and oblate configurations. Our results predict deformed shapes for a
large number of isotopes for all the three chains.  The deformations
of nuclei have been obtained from the relativistic Hartree
minimization. We show in Fig. 6 the quadrupole deformation $\beta_2$
for the shape corresponding to the lowest energy. The predictions of
FRDM and ETF-SI are also shown for comparison. In all the three isotopic
chains, the RMF theory gives a well-defined prolate shape for lighter
isotopes. Further, an addition of a few neutrons below the closed neutron
shell leads to an oblate shape in all the three cases. These shapes
turn into spherical ones as nuclei approach the magic neutron number
N=50. Nuclei above this magic number revert again to the prolate shape
in the RMF theory. Thus, in all the three isotopic chains a shape transition
from prolate-oblate-spherical-prolate is followed. Only in the Kr isotopes,
the very heavy Kr nuclei assume an oblate shape. The very neutron-rich
Sr and Zr nuclei, on the other hand, retain a highly deformed prolate
shape.

In addition to the lowest minimum, several isotopes exhibit a second
minimum, thus implying a shape-coexistence, i.e., the prolate and the
oblate shapes differ in the energy only by a few hundred keV. These nuclei
have been shown by squares surrounding the black circles.
Our calculations predict two minima for several heavy Sr nuclei,
the prolate shape results being a few hundreds keV lower in
energy than the oblate one. This is displayed in Fig. 7, where the
difference in the ground-state binding energy of the oblate and the
prolate configurations is shown. It can be noticed that Sr isotopes
beginning with A=92 acquire a prolate shape predominantly. For nuclei
close to A=98, the prolate shape is lower than the oblate one only
by about 300 keV. With a further increase in the neutron number the
Sr isotopes take up the prolate minimum, the oblate shape being
about 600-800 keV higher. The undulation in the prolate-oblate energy
differences of neutron-rich Sr isotopes is a noteworthy feature of
the RMF prediction.

As seen from Fig. 6 (b), most of the Sr isotopes acquire a deformed
shape except those close to the neutron number N=50. This is in marked
contrast with earlier calculations \cite{maha92} using the force NL1,
where spherical shapes were obtained for a large number of Sr nuclei.
This has also been the case for many Zr isotopes in a previous
work \cite{shei93} with the force NL1 where spherical shapes were
predicted. This is significantly different from the results of the
Zr isotopes with the force NL-SH as shown in Fig. 6 (c),
where large deformations and shape-transitions have been observed.
Experimentally, nuclei in this region are susceptible to
gaining very large deformations. Sr nuclei are a good case of such a
property and the onset of deformation in nuclei close to A=100
is now well known \cite{conf}. The strong empirical onset of deformation
in the Mo chain \cite{raman87} is being exhibited also by our Sr and
Zr results with the force NL-SH.

A comparison of our results with those of the FRDM shows that
by and large there is a reasonable agreement between the $\beta_2$ values
of NL-SH and the predictions of FRDM. The striking difference between
the two is that a shape transition from prolate to oblate in the lighter
Sr and Zr nuclei in the RMF theory is not given by the FRDM. The FRDM,
in contrast, provides a shape transition from prolate to about spherical
on approaching the magic neutron number in these two chains. Nuclei
heavier than semi-magic Sr and Zr isotopes are predicted to take up
highly deformed prolate shapes in the FRDM. This is consistent with our
results using NL-SH. The scenario with the Kr chain is slightly different
from the other two. In the beginning of this chain at A=72, the RMF as well
as FRDM show an additional shape transition from oblate to prolate. The
shape transition from prolate-oblate-spherical-prolate is provided in
both the RMF and the FRDM. Only for very heavy Kr isotopes, nuclei
in the FRDM assume a highly deformed prolate shape, as against the oblate
shape predicted by the RMF.

The ETF-SI model \cite{pear93} also predicts deformation properties
of a large number of nuclei all over the periodic table. The $\beta_2$
values for all the three isotopic chains are shown in Fig. 6. A peculiar
feature of the ETF-SI predictions is that on going from light nuclei
to the heavier ones, one encounters a shape transition from prolate to oblate,
oblate to spherical, and then again spherical to oblate for both the Sr and
Zr isotopic chains. The latter shape transition from  spherical to oblate,
which is predicted by ETF-SI in all the three chains, is not
supported by our results nor by the predictions of the FRDM.
For heavier nuclei, ETF-SI shows a transition from oblate to prolate
directly without going first to a spherical shape. Moreover, for the
very light Kr isotopes, ETF-SI provides only a very deformed oblate
shape, in contrast with the results of RMF and FRDM. However,
shape transition from prolate to oblate in the Kr chain at A=92
is consistent with the one obtained in the RMF theory. At this point,
the FRDM predicts only a prolate shape for the heavy Kr nuclei.
Thus, the significant discrepancy between the ETF-SI on one hand and
RMF and FRDM on the other hand, lies in the shape transition from prolate
to oblate for heavier isotopes in all the three chains in ETF-SI, which is
not present in the other two approaches.

The $\beta_2$ values from NL-SH, FRDM and ETF-SI are shown in Tables 4-6.
The $\beta_2$ values for the second minimum obtained for Kr and Sr isotopes
with the force NL-SH are shown in the parentheses. A comparison of the
magnitudes of the deformation for all the three isotopic chains shows that
the three approaches provide values close to each other. In addition,
we have also performed calculations with the RMF force NL1 for Sr and Zr
isotopes. The resulting $\beta_2$ values are shown in Tables 5 and 6.
Futhermore, we also include for comparison $\beta_2$ values available for
Zr isotopes using the Skyrme force SIII and taken from \cite{bonche85}.
The experimental quadrupole deformations obtained from BE(2) values taken
from \cite{raman87} are shown in the last columns of the tables. It may be
noted that these empirical $\beta_2$ values do not indicate any signature
as to ascertain the shape of a given nucleus. The absolute values, however,
do compare well with the RMF predictions.

\bigskip
\section{\sf SUMMARY AND CONCLUSIONS}
The RMF theory has been employed to obtain the ground-state properties
of the isotopic chains of Kr, Sr and Zr nuclei. This region of nuclei
which entails several dramatic transformations of shapes
has not been amenable to an appropriate description in the conventional
approaches such as the density-dependent Skyrme Hartree-Fock theory.
Nuclei in this region exhibit anomalous behaviour in the isotope shifts,
resulting in rms charge radii of lighter isotopes being larger than the
heavier closed neutron-shell counterparts with N=50. This is a generic
feature of many isotopic chains in this region, and is attributable
primarily to the shape transitions very common in this region.
The RMF theory with the force NL-SH renders a successful
description of the general behaviour of the isotopic shifts in Kr
and Sr chains. It also predicts a similar trend for Zr chain where
empirical data do not yet exist and measurements are required.
It is noteworthy that for the first time a microscopic theory
is able to describe the empirical data on the isotope shifts.
This success of the RMF theory should be viewed in conjunction
with its ability to describe \cite{sha93b} also the kink in the isotope
shifts across shell-closure in Pb nuclei.

A non-relativistic reduction of the RMF Hamiltonian shows \cite{koen94}
that the isospin dependence of the spin-orbit interaction which stems
naturally in the RMF theory as a consequence of the Lorentz covariance
and Dirac structure of nucleons is not contained in the density-dependent
Skyrme theory. Consequently all attempts in the Skyrme theory
including a plausible influence of the ground-state correlations have
failed to explain the anomalous kinks in the isotope shifts of Pb nuclei
and nuclei in the Z=40 region.

Deformation properties of nuclei are pivotal to
the successful description of the isotope shifts.
The deformations which bring about the anomalous isotope shifts, are
generally in good agreement with the $\beta_2$ values obtained from
the BE(2) measurements.  The RMF theory predicts prolate-oblate shape
coexistence in heavy Sr nuclei. The shape-coexistence in a few Kr isotopes
has also been shown. The RMF theory also predicts several shape transitions
along the mass chains. Many of these shape changes are contained in either
one or the other of the FRDM and ETF-SI mass formulae. ETF-SI deformations
in many cases are just opposite to the predictions of the RMF and the FRDM.
This fact is consequently reflected in the difficulty of the Skyrme approach
to describe the isotope shifts on Sr and Kr nuclei.

The binding energies of nuclei in all the chains considered here
have been described successfully in the RMF theory. The deviations
in the binding energies from the empirical values are at the most
0.5\%. The parabolic behaviour of the binding energy per nucleon
has been obtained for all the chains. This is consistent with the
empirical curves and also with the extensive mass fits of FRDM and
ETF-SI. The agreement manifests the degree of the success of the
RMF predictions even with a force obtained from a limited adjustment
only on 6 nuclei. In conjunction with good description of
the deformations and isotope shifts, the RMF theory provides a
unified description of the ground-state propertie of nuclei over
a large range of isospin.

\bigskip
\section{\sf Aknowledgment}
Fruitful discussions with Prof. E.W. Otten and Dr. R. Neugart are thankfully
acknowledged. We thank Peter Ring for useful discussions. Thanks are also
due to Prof. Brian Serot for pointing out an inadvertant error.
One of the authors
(G.A.L) aknowledges support by the European Union under the contract
HCM-EG/ERB CHBICT-930651.


\newpage
\noindent\begin{table}
\begin{center}
\caption{\sf The binding energy (MeV) of Kr isotopes obtained with the force
NL-SH. The predictions from the mass models FRDM {\protect\cite{nix94}} and
ETF-SI {\protect\cite{pear93}} are also shown for comparison. A comparison has
also been made with the binding energies in the Skyrme approach
{\protect\cite{bonche85}} with the force SIII, wherever available.
The empirical values {\protect\cite{audi93}} are shown in the last column.}
\bigskip
\begin{tabular}{ll c c c c c c l}
\hline\hline
& A & NL-SH & FRDM & ETF-SI & SIII & expt.&\\
\hline
&70& 575.87& 578.33& 577.28& -& 577.80 &\\
&72& 604.00& 607.00& 605.02& -& 607.11 &\\
&74& 628.49& 631.99& 630.13& 626.42& 631.28 &\\
&76& 651.64& 654.82& 653.56& 648.90 & 654.23 &\\
&78& 672.69& 675.56& 675.23&- & 675.55 &\\
&80& 693.45& 695.05& 695.35&- & 695.43 &\\
&82& 713.39& 714.57& 714.26&- & 714.27 &\\
&84& 733.16& 732.69& 732.12&- & 732.26 &\\
&86& 750.06& 748.97& 748.79&- & 749.23 &\\
&88& 760.08& 761.22& 761.21&- & 761.80 &\\
&90& 769.89& 771.95& 772.60& 769.42 & 773.21 &\\
&92& 779.84& 782.35& 782.25& -& 783.22 &\\
&94& 789.88& 791.89& 791.18& -& 791.76&\\
&96& 799.46& 800.85& 799.33& -& 799.95&\\
&98& 807.51& 808.87& 806.79& 804.64& -&\\
&100& 814.02& 815.68& 813.37& 811.44& -&\\
\hline\hline
\end{tabular}
\end{center}
\end{table}
\vfill
\noindent\begin{table}
\begin{center}
\caption{\sf The binding energies for Sr isotopes. See the caption of Table 1
for details. In addition, the binding energies obtained in the RMF with the
force NL1 are also shown.}
\bigskip
\begin{tabular}{ll c c c c c c l}
\hline\hline
&A& NL-SH& NL1& FRDM& ETF-SI& SIII& expt.&\\
\hline
&70& 542.39&  -   &543.98 &545.40&   -   &   -   &\\
&72& 574.69&579.51&577.48 &577.70&   -   &   -   &\\
&74& 606.74&608.59&609.17 &607.99&603.97 &   -   &\\
&76& 635.95&638.35&638.66 &636.47&633.56 & 638.08&\\
&78& 661.71&663.40&663.90 &662.13&658.53 & 663.01&\\
&80& 683.76&688.35&684.82 &685.56&680.74 & 686.28&\\
&82& 705.86&710.03&707.62 &707.58&   -   & 708.13&\\
&84& 726.79&730.51&729.15 &728.51&   -   & 728.90&\\
&86& 748.81&750.54&749.43 &748.73&747.70 &748.92&\\
&88& 767.96&768.84&768.09 &767.88&768.99 &768.46&\\
&90& 779.76&780.38&781.81 &781.74&780.35 &782.63&\\
&92& 792.05&791.60&794.07 &794.85&791.05 &795.75&\\
&94& 803.83&802.87&806.53 &806.34&802.23 &807.81&\\
&96& 814.87&813.74&818.13 &817.16&813.41 &818.10&\\
&98& 826.60&823.08&828.98 &827.82&825.45 &827.87&\\
&100& 835.63&832.09&838.70&837.22&834.77 &837.62&\\
&102& 844.71&841.50&847.24&845.57&843.42 &846.62&\\
&104& 852.56&848.80&855.14&852.79&  -    &  -   &\\
&106& 859.77&  -   &862.58&859.22&  -    &  -   &\\
&108& 865.61&  -   &868.78&864.69&  -    &  -   &\\
&110& 870.50&  -   &873.46&869.45&  -    &  -   &\\
\hline\hline
\end{tabular}
\end{center}
\end{table}
\vfill

\newpage
\noindent\begin{table}
\begin{center}
\caption{\sf The binding energies for Zr isotopes. See the caption of Table 2
for details.}
\bigskip
\begin{tabular}{ll c c c c c c l}
\hline\hline
&A& NL-SH&NL1& FRDM& ETF-SI& SIII& expt.&\\
\hline
&80&667.20&668.85&669.26&666.99&663.73&669.75&\\
&82&691.39&697.52&692.30&692.53&688.30&694.74&\\
&84&715.79&720.72&717.23&716.89&713.11&718.19&\\
&86&738.91&744.11&740.80&740.38&737.44&740.65&\\
&88&762.02&766.09&763.18&762.78&761.21&762.61&\\
&90&784.42&785.69&784.11&783.94&784.41&783.89&\\
&92&797.19&799.91&799.43&799.55&797.51&799.72&\\
&94&811.62&812.16&813.42&813.94&810.04&814.68&\\
&96&825.43&824.64&827.25&827.51&823.08&828.99&\\
&98&838.07&835.94&840.74&840.10&835.88&840.96&\\
&100&851.40&847.98&853.37&852.43&849.25&852.44&\\
&102&862.83&858.58&864.82&863.70&860.23&863.72&\\
&104&872.79&868.44&875.02&873.82&870.61&874.46&\\
&106&882.27&876.62&884.43&882.76&  -   &884.45&\\
&108&891.00&884.68&893.38&890.75&  -   &  -  &\\
&110&898.28&890.74&901.17&897.90&  -   &  -  &\\
\hline\hline
\end{tabular}
\end{center}
\end{table}
\newpage
\noindent\begin{table}
\begin{center}
\caption{\sf The quadrupole deformations $\beta_2$ for Kr isotopes
obtained in the RMF theory using the force NL-SH. The FRDM and ETF-SI
predictions are also shown. The available empirical deformations (expt.)
obtained from BE(2) values are also given in the last column. The experimental
values do not depict the signature of the deformation. The deformations of
nuclei showing a shape coexistence with a second minimum are given in
the parentheses.}
\bigskip
\begin{tabular}{lll c  c c c cl}
\hline\hline
& A & NL-SH  & FRDM& ETF-SI& expt.& \\
\hline
&72 & -0.298&-0.349&-0.40&  -&\\
&74& 0.441(-0.318)&0.400&-0.30&0.387 &\\
&76&0.431(-0.267)&0.400&-0.30& 0.408& \\
&78&-0.222(0.339)&-0.232&-0.30& 0.343&\\
&80&-0.204&0.002&-0.30& 0.265 &\\
&82&0.110&0.071&0.18& 0.202 &\\
&84&0.0&0.062&0.15& 0.149 &\\
&86&0.0&0.053&-0.04& 0.145&\\
&88&0.049&0.062&-0.13&-&\\
&90&0.175(-0.175)&0.162&0.15& -&\\
&92&0.228(-0.224)&0.228&-0.21&-&\\
&94&-0.264&0.310&-0.29&-&\\
&96&-0.298&0.337&-0.31&-&\\
&98&-0.292&0.349&0.36&-&\\
&100&-0.284&0.350&0.38&-&\\
\hline\hline
\end{tabular}
\end{center}
\end{table}
\newpage
\noindent\begin{table}
\begin{center}
\caption{\sf The quadrupole deformations $\beta_2$ for Sr isotopes. See
Table 4 for details. The values obtained from the RMF force NL1 are also
shown here.}
\bigskip
\begin{tabular}{lll c c c c c cl}
\hline\hline
& A & NL-SH & NL1 &FRDM&ETF-SI& expt.& \\
\hline
&72 & 0.324&-0.205&0.371&-0.30   &  -&\\
&74& 0.430&0.432&0.400&0.44  &- &\\
&76&0.450&0.464&0.421&0.44  & -& \\
&78&0.450&0.456&0.421& 0.43 & 0.434&\\
&80&0.402&0.0&0.053&0.40  & 0.377 &\\
&82&-0.200&0.0&0.053&-0.30 & 0.290 &\\
&84&0.089&0.0&0.053&0.15  & 0.211 &\\
&86&0.0&0.0&0.053&0.00 & 0.128&\\
&88&0.0&0.0&0.045&0.00 & 0.117&\\
&90&0.0&0.0&0.045&-0.11 & -&\\
&92&0.181(-0.165)&0.097&0.080&-0.15 &-&\\
&94&0.230(-0.218)&-0.187(0.180)&0.255&-0.19 &-&\\
&96&0.356(-0.275)&-0.242(0.207)&0.338&0.35 &-&\\
&98&0.424(-0.309)&-0.218(0.459)&0.357&0.39 &0.354&\\
&100&0.426(-0.314)&-0.194(0.439)&0.368&0.38 &0.372&\\
&102&0.413(-0.295)&-0.172&0.369&0.40 & -&\\
&104&0.403(-0.277)&-0.139&0.361&0.38 & -&\\
\hline\hline
\end{tabular}
\end{center}
\end{table}
\newpage
\noindent\begin{table}
\begin{center}
\caption{\sf The quadrupole deformations $\beta_2$ for Zr isotopes. See
Table 5 for details. The values from the Skyrme force SIII are also given.}
\bigskip
\begin{tabular}{lll c c c c c c c l}
\hline\hline
& A & NL-SH & NL1 & SIII&FRDM&ETF-SI& expt.& \\
\hline
&80&0.463&0.466&0.490&0.433&0.45&- &\\
&82&0.429&0.0&-0.210&0.053&0.41&- &\\
&84&-0.201&0.0&0.0&0.053&-0.26& 0.250 &\\
&86&-0.145&0.0&0.0&0.053&0.00& 0.149&\\
&88&0.0&0.0&0.0&0.053&0.00& 0.187&\\
&90&0.0&0.0&0.0&0.035&0.00& 0.091&\\
&92&0.099&0.0&0.0&0.053&0.00&0.102&\\
&94&0.197&0.0&0.232&0.062&0.02&0.090&\\
&96&0.243&0.0&0.232&0.217&-0.19&0.081&\\
&98&0.337&0.0&0.344&0.330&-0.27&-&\\
&100&0.397&0.453&0.425&0.358&0.36&0.321&\\
&102&0.404&0.427& 0.438&0.369&0.380&0.421&\\
&104&0.402&0.418& 0.440&0.381&0.40&-&\\
\hline\hline
\end{tabular}
\end{center}
\end{table}

\newpage
\leftline{\bf Figure Captions.}
\bigskip
\bigskip
{\bf Fig. 1} The binding energy per nucleon for Kr, Sr and Zr isotopes
in the RMF theory with the force NL-SH. The predictions from the mass
models FRDM and ETF-SI are also shown for comparison. The RMF binding
energies exhibit a clear parabolic behaviour with the mass number and
compare reasonably well with the empirical data (expt).

\bigskip

{\bf Fig. 2} The isotope shifts for Kr, Sr and Zr nuclei calculated with
the force NL-SH. The experimental isotope shifts for Kr \cite{otto94}
and Sr \cite{ott88,buch90} nuclei are also shown. The prediction for Zr
isotope shifts is very similar to the behaviour of the Kr and Sr nuclei

\bigskip

{\bf Fig. 3.} The rms charge and neutron radii of Kr, Sr and Zr isotopic
chains obtained in the RMF theory using the force NL-SH.
\bigskip

{\bf Fig. 4.} The isotope shifts for Zr nuclei in the RMF theory compared
with the prediction of the Skyrme force SIII.

\bigskip

{\bf Fig. 5.} The neutron-skin thickness of Zr isotopes in the RMF theory
with the forces NL-SH and NL1. The NL1 values are higher than those of NL-SH
due to the correspondingly high asymmetry energy of NL1 as compared to
that of NL-SH and the empirical value. The predictions of the Skyrme force
SIII are even lower than the NL-SH values (see text for details).

\bigskip

{\bf Fig. 6.} The quadrupole deformation $\beta_2$ obtained from relativistic
Hartree minimization for Kr, Sr and Zr isotopes using the force NL-SH. The
predictions of the mass models FRDM and ETF-SI are displayed for comparison.
Nuclei exhibiting a shape coexistence and  thus a second minimum
in the RMF theory are depicted by a square surrounding the $\beta_2$ value
of the lowest minimum. Several shape transitions in all the three isotopic
chains emerge as the characteristic feature of the deformation properties in
the RMF theory.

\bigskip

{\bf Fig. 7.} The prolate-oblate shape coexistence for neutron-rich Sr
isotopes predicted in the RMF theory. The energy difference in the prolate
and oblate minima for Sr isotopes is shown. Most of the nuclei in the
neutron-rich domain possess a prolate minimum. An oblate second minimum
lies only a few hundred keV higher in energy.

\vfill

\end{document}